# Low-frequency Current Fluctuations in Individual Semiconducting Single-Wall Carbon Nanotubes

Yu-Ming Lin[1],[*] Joerg Appenzeller[1], Joachin Knoch[2], Zhihong Chen[1], and Phaedon Avouris[1]

[1]*IBM T. J. Watson Research Center,*
*Yorktown Heights, NY 10598, USA*

[2]*Institute of Thin Films and Interfaces,*
*Forschungszentrum Julich, Julich 52454, Germany*

## Abstract

We present a systematic study on low-frequency current fluctuations of nano-devices consisting of one single semiconducting nanotube, which exhibit significant $1/f$-type noise. By examining devices with different switching mechanisms, carrier types (electrons vs. holes), and channel lengths, we show that the $1/f$ fluctuation level in semiconducting nanotubes is correlated to the total number of transport carriers present in the system. However, the $1/f$ noise level per carrier is not larger than that of most bulk conventional semiconductors, e.g. Si. The pronounced noise level observed in nanotube devices simply reflects on the small number of carriers involved in transport. These results not only provide the basis to quantify the noise behavior in a one-dimensional transport system, but also suggest a valuable way to characterize low-dimensional nanostructures based on the $1/f$ fluctuation phenomenon.

# INTRODUCTION

Carbon nanotubes and nanowires are promising candidates for advanced nanoelectronic devices, and they have demonstrated great potential in a wide range of applications, such as field-effect transistors[1–3], elementary logic circuits[4–6], and chemical sensors[7, 8]. However, nanotube devices have been shown to exhibit very significant current fluctuations in the low-frequency regime[9], which may present serious limitations for device performance and applicability, e.g. the threshold voltage of a transistor and the detection sensitivity of a chemical sensor. Despite the broad interest in nanotubes and other one-dimensional systems, a comprehensive study on their noise characteristics is still lacking. On the other hand, while the noise is usually regarded as an undesirable property for applications, the fluctuation phenomenon in itself contains important information about the material and may be utilized as a valuable probe to characterize nanostructures.

In solid-state electronic devices, two types of fluctuations usually dominate the low-frequency noise behavior at room temperature: (a) thermal noise associated with statistical Brownian motion [10] and (b) the so-called $1/f$ noise that exhibits a power spectrum varying inversely proportional to the frequency $f$. The $1/f$ noise typically involves fluctuation processes with a distribution of time constants, and is, in fact, a ubiquitous type of fluctuation, appearing not only in electronic devices but also in a wide variety of physical systems, including biological and geological systems. Other types of noises, such as shot noise, generation-recombination noise, and random telegraph noise, may also be present in solid-state devices and become important under certain temperature and bias conditions.

Previous experiments on the noise properties of carbon nanotube-based devices have revealed a pronounced $1/f$ noise component in the low-frequency regime [9, 11, 12]. It has been speculated that carbon nanotubes always possess a $1/f$ noise much worse than conventional bulk materials. However, attempts to further examine and analyze the origin of noise in carbon nanotubes were hampered by the coexistence of various nanotubes in the form of 2D films or 3D networks used in previous studies. Here we perform, for the first time, a systematic study on $1/f$ fluctuations of carbon nanotube field-effect transistors (CNFETs) consisting of *individual* semiconducting single-wall nanotubes, and show that the pronounced $1/f$ noise in carbon nanotubes is associated with the small number of transport carriers in the device.

**DEVICE LAYOUT**

Semiconducting nanotubes are used in our study because of their technological potential as active electronic components as well as the ability to control their device characteristics by the gate voltage and/or dopants. CNFETs with two different gate structures are fabricated and investigated in order to distinguish between the role of the contacts and bulk channel. The first type of CNFET, denoted as SB-CNFET (Fig. 1(a)), is a back-gated device where the device conductance is entirely modulated by Schottky barriers (SBs) at the nanotube/metal interfaces[13, 14]. The second type of CNFET incorporates an additional gate electrode so that the device switching can be achieved through the bulk (middle) part of the nanotube[15] (Fig. 1(b)), which we denote as C-CNFET (channel switching). We have also fabricated SB-CNFETs with very different channel lengths out of a *single* long nanotube (Fig. 1(c)) in order to elucidate the significance of scattering on $1/f$ noise. In spite of distinctly different switching mechanisms, we find that both $p$-, $n$-type SB-CNFETs and C-CNFETs all exhibit a $1/f$ noise level that increases with device resistance $R$ as the gate voltage is varied. For SB-CNFETs with different channel lengths, we observe a $1/f$ noise level that is inversely proportional to the device channel length. Overall, these results show that (i) the magnitude of $1/f$ noise in a 1D conductor is correlated to the total number of free carriers in the system; (ii) both fluctuations occurring at the contacts and along the nanotube channel contribute to the $1/f$ noise phenomenon in CNFETs; (iii) the fluctuation mechanism is independent of carrier (electron or hole) type. Most importantly, the intrinsic $1/f$ noise amplitude of individual single-wall carbon nanotubes is, contrary to previous suggestions[9], not larger than that of most conventional bulk materials. While the $1/f$ noise in bulk materials has been proposed to be induced by the scattering with surface or/and bulk phonons[16], we find that the noise amplitude in carbon nanotubes is not affected by the presence of acoustic phonon scattering or ionized impurity scattering.

**SCHOTTKY BARRIER-DOMINATED CNFET**

Fig. 2(a) shows the subthreshold characteristics of an SB-CNFET, exhibiting regular $p$-type FET behavior where the drain current $I_d$ decreases with increasing gate voltage $V_g$. Pronounced current fluctuations are clearly observed for a given gate and drain voltage.

We note that on a logarithmic scale, the vertical spreading of current data points is a measure of the relative current fluctuation $\Delta I_d/I_d$. Although random telegraph signals (RTSs) have been observed in some carbon nanotube devices at low temperatures [17], no RTS-like features are present in our CNFETs at room temperature (see Fig. 2(b)). As shown in Fig. 2(b), at a constant $V_d = -0.5\,\text{V}$, the relative current fluctuation $\Delta I_d/I_d$ of the CNFET rises significantly with increasing (more positive) gate voltage. However, $\Delta I_d/I_d$ shows little dependence, if any, on the absolute current level and the drain voltage. This indicates that the observed current fluctuation $\Delta I_d$ is not due to the thermal noise which is current independent (i.e. $\Delta I/I \propto I^{-1}$) or the shot noise which is proportional to $I^{0.5}$ (i.e. $\Delta I/I \propto I^{-0.5}$) [18]. In fact, we find that the current fluctuation $\Delta I_d$ is proportional to $I_d$ and the ratio $\Delta I_d/I_d$ only depends on the gate voltage.

Fig. 2(c) shows the noise power spectrum $S_I$, normalized by $I_d^2$, of the SB-CNFET at various gate voltages. At a finite drain bias, the power spectral density varies as $f^{-\beta}$, where $\beta$ is close to 1 within 10%, and is thus referred to as the $1/f$ noise. Similar to other electronic devices that exhibit $1/f$ noise, including the carbon nanotube-based devices previously studied [9], the excess $1/f$ noise power of CNFETs is proportional to the square of the dc current $I^2$, which can be expressed as

$$\frac{S_I}{I^2} = \frac{S_R}{R^2} = \frac{A}{f^\beta} \qquad (1)$$

in the linear resistance regime, where $A$ is defined as the $1/f$ noise amplitude and $\beta \simeq 1$. In Fig. 2(c), the noise level increases by almost two orders of magnitude as the the gate voltage $V_g$ increases from -0.6 V to 0.6 V, consistent with the trend of $\Delta I_d/I_d$ at different $V_g$ shown in Fig. 2(b). In Fig. 3(a), we plot the drain current and the $1/f$ noise amplitude $A$ of the CNFET as a function of gate voltage. As the device current diminishes with increasing $V_g$, the noise amplitude $A$ increases monotonically with the device resistance $R = V_d/I_d$, with an $A/R$ ratio that lies between $2 \times 10^{-10}$ and $2 \times 10^{-9}\,\Omega^{-1}$.

Based on the statistics of a large number of devices consisting of mixtures of different types of carbon nanotubes (i.e. 2D mats and 3D networks), it has been concluded that the $1/f$ noise amplitude in these carbon nanotube-based devices increases with the sample resistance $R$ [9], with an $A/R$ ratio depending on device dimensions [11]. It was, however, not possible to study the origin of this correlation between $A$ and $R$ in those experiments because of considerable sample-to-sample variations and a wide, uncontrolled distribution of nanotubes even in a single device. Since our gate-dependent noise measurements are performed on the *same* CNFET device consisting of a *single* nanotube, the uncertainty due to device-to-device variations is eliminated. The well-defined transport channel also allows for quantitative simulations to study the $1/f$ noise phenomenon in an ideal one-dimensional semiconductor.

To understand the gate-dependent noise amplitude in Fig. 3(a), we first examine the CNFET operation mechanisms and consider factors that vary with the gate voltage $V_g$. In an SB-CNFET, important $V_g$-dependent device attributes relevant to transport phenomena include the number of carriers within the nanotube channel and the SB profile at the contacts. Since the carrier mean-free-path can be as long as $\mu$m in nanotubes at low fields [19], the carrier transport is quasi-ballistic for a channel length of 600 nm in our SB-CNFET, and the device resistance is mainly determined by the SB at the contacts [13, 14]. For the $p$-type CNFET shown in Fig. 2(a), the CNFET is switched ON at sufficiently negative gate voltages because the Schottky barrier at the valence band is effectively thinned to enable hole carrier injection via direct tunnelling. At the same time, the number of hole carriers in the nanotube channel also increases with decreasing (more negative) $V_g$. Although the resistance of an SB-CNFET may be exclusively modulated by the SB width, the noise measurements of C-CNFETs (see discussion below and Fig. 4) indicate that the $1/f$ noise variation in a carbon nanotube can not be fully accounted for by the gate-dependent SB profile at the contact alone.

We demonstrate below that the gate-dependent $1/f$ noise in semiconducting nanotubes is governed by the total number of transport carriers $N$ in the channel. We find that the noise amplitude $A$ is inversely proportional to $N$, expressed by

$$A = \frac{\alpha_H}{N}, \tag{2}$$

where $N$ is the number of carriers in the system and $\alpha_H$ is a parameter that may vary from device to device. This $A \propto 1/N$ dependence was first introduced by Hooge [20] to describe the $1/f$ noise results in homogenous bulk materials with an empirical value of $\alpha_H \simeq 2 \times 10^{-3}$. Although $\alpha_H$ is not a universal constant, $\alpha_H \sim 10^{-3}$ is a value frequently found in non-optimized bulk materials [21]. While Eq. (2) holds true for most bulk and thin film specimens with diffusive transport behaviors, there has been no experimental verification of the relation and/or information about $\alpha_H$ for one-dimensional systems exhibiting quasi-ballistic transport behavior. One of the obstacles is that the number of carriers in such one-dimensional objects cannot be readily measured by conventional techniques such as Hall effect or Shubnikov-de Haas oscillation measurements. This carrier number information, however, may be obtained from a reliable transport model that properly reproduces experimental results. In this respect, the back-gated CNFET (Fig. 1(a)) is instrumental because both the simplicity of the device geometry and the well-established band-structure in carbon nanotubes facilitate dependable model simulations.

Theoretical modeling of electrical properties of SB-CNFETs has been extensively developed among different groups [22–24], with good agreement obtained between simulation and experimental data [25, 26]. Here we adopt the model described in Ref. [23], which is based on the non-equilibrium Green's function formalism[27], and calculate the device current at $V_d = 10\,\text{mV}$ as a function of gate voltage (solid curve in Fig. 3(a)). The simulated result exhibits satisfactory agreement with experimental data in terms of both the absolute current level and the gate dependence. With the success of reproducing the CNFET current, the same model is then employed to calculate the total number of transport carriers $N$ by integrating the hole carrier density in the valence band over the nanotube channel between two metal contacts for the given bias conditions [28]. For a channel length of 600 nm, the model predicts a total number of holes $N \sim 40$ in the ON state[29], and $N$ decreases as the CNFET is switched OFF with increasing gate voltage. It should be noted that because of the finite transmission probability at the nanotube/metal contacts and the associated delocalized nature of the electron/hole wave functions, the calculated carrier number $N$ is not necessarily an integer and can be smaller than unity near the OFF state. In Fig. 3(a), we plot the simulated noise amplitude $A$ (dashed curve) based on Eq. (2) when $\alpha_H = 2 \times 10^{-3}$, showing excellent agreement between the measurements and the simulation.

We also performed noise measurements on $n$-type SB-CNFETs, which were fabricated by exposing the device to potassium (K) vapor [30], and measured gate-dependent currents $I_d$ and noise amplitudes $A$ that mirror their $p$-type counterparts (see Fig. 3(b)). These results present strong evidence that the gate-dependent $1/f$ noise observed in carbon nanotubes is modulated by the total number of transport carriers in the channel according to Eq. (2), and the fluctuation mechanism is independent of the carrier type, i.e. electrons or holes. Furthermore, the $1/f$ noise parameter $\alpha_H = 2 \times 10^{-3}$ determined for the CNFETs that consist of unpurified carbon nanotubes is quite comparable to the value observed in most bulk systems. The much larger $\alpha_H$ ($\sim 0.2$) deduced in previous carbon nanotube devices [9] is due to an overestimation of $N$ by adopting the number of atoms instead of the number of transport carriers, as well as the lack of a quantitative model to accurately extract $N$. The fact that the pronounced $1/f$ noise amplitude $A$ observed in carbon nanotubes is due to a much smaller $N$ value rather than a material-specific property indicates that this elevated noise level is a general property of nanoscale electronic devices. Since the $1/N$ dependence and mechanisms for the $1/f$ noise phenomenon are, to the lowest order, independent of the system dimensionality, various techniques that have been developed for reducing the $1/f$ noise in bulk systems may also be applicable for nanotubes and other 1D nanostructures.

**CHANNEL-LIMITED CNFET**

In order to clarify the role of the SB contacts on the $1/f$ noise of a CNFET, we have fabricated dual-gate CNFETs [15], called C-CNFETs, where an additional Al gate stack is placed underneath the nanotube between the source/drain electrodes (Fig. 4(a)). When the Si backgate is kept at a constant negative (or positive) voltage, the transistor operation is achieved by the Al gate voltage that alters the potential barrier within the nanotube channel in a way similar to that of conventional MOSFETs. With the dual-gate structure, the band-bending at the contacts and in the nanotube channel can be independently controlled, making it possible to distinguish their roles in the noise behavior.

As shown in Fig. 4(b), the C-CNFET operates as a $p$-type FET at a negative Si backgate voltage. Similar to the SB-CNFET, the low-frequency current fluctuation of the C-CNFET is also dominated by the $1/f$ noise, and $S_I$ is proportional to $I_d^2$. Fig. 4(b) shows that the noise amplitude $A$, derived according to Eq. (1), increases with the resistance as $V_{g-Al}$ varies, with an $A/R$ ratio between $3\times 10^{-10}$ and $3\times 10^{-9}\,\Omega^{-1}$. We note that since the SB potential profile in C-CNFETs is unchanged at a constant $V_{g-Si}$, this significant variation of $A$ as a function of $V_{g-Al}$ can only be accounted for by the proposed $N$-dependent $1/f$ noise behavior. We also observe that, for the same Al gate voltage, the C-CNFET possesses a higher $1/f$ noise level at $V_{g-Si} = -1.5\,\text{V}$ compared to that at $V_{g-Si} = -2.5\,\text{V}$ (see Fig. 4(b)). This is because the number of carriers for a given $V_{g-Al}$ is smaller due to the lower transmission probability through the SB modulated by $V_{g-Si}$. The results of the C-CNFET for different Si and Al gate configurations illustrate that the $1/f$ noise level in a semiconducting nanotube cannot be solely specified by either the contact or channel characteristics. Interestingly, despite the different switching mechanisms of an SB-CNFET and a C-CNFET, both devices exhibit very similar dependence of the $1/f$ noise amplitude on resistance. Overall, these observations strongly suggest that the variation of $A$ in both types of devices originates from a common factor that is related to a fundamental property of the entire device, i.e. the total number of transport carriers $N$. The $1/f$ noise amplitude rises with decreasing $N$, regardless of whether $N$ is changed by altering the injection into the nanotube channel or by introducing a barrier inside the nanotube as in the case of a C-CNFET. In both type of structures (SB- and C-CNFET), the current fluctuations $\Delta I_d$ are caused by surface potential variations $\Delta \phi$ that modulate the transmission probability at the contact and/or across the channel. The surface potential fluctuations $\Delta \phi$ are due to the number fluctuation $\Delta N$ of the transport carriers in the channel, which may result from trapping-detrapping processes in the oxide, or motion of adsorbed species on or inside the nanotube. Although the fluctuation process may depend on the structure of contacts and the quality of oxide, the potential fluctuation $\Delta \phi$ always becomes more pronounced as $N$ decreases (e.g. as gate voltage varies). Further studies, however, are necessary to distinguish and identify the origin of various fluctuation processes in a CNFET device.

## CHANNEL LENGTH DEPENDENCE

To further demonstrate the $N$ dependence and to evaluate the significance of scattering on the $1/f$ noise of nanotubes, we have fabricated two SB-CNFETs with very different channel lengths: $l \simeq 500\,\mathrm{nm}$ and $7\,\mu\mathrm{m}$ (inset of Fig. 5), which are denoted as short ($S$) and long ($L$) devices, respectively. To eliminate device-to-device variations, the two CNFETs are fabricated using one, single, long semiconducting nanotube and share a common electrode as the source. Both devices possess regular $p$-FET characteristics and exhibit a similar gate voltage dependence, with low-bias ON-state resistances (at the same gate voltage) of $\simeq 10\,\mathrm{k\Omega}$ and $58\,\mathrm{k\Omega}$ for the short and the long devices, respectively. Fig. 5 plots the ON-state noise power spectrum measured at $V_d$=10 mV, showing the $1/f$ noise with amplitudes $A_S \sim 4 \times 10^{-6}$ and $A_L \sim 4 \times 10^{-7}$ for the short and the long CNFETs, respectively. In spite of the higher resistance of the long device, its $1/f$ noise level is about an order of magnitude lower than that of the short device. We note that the ratio of the noise amplitude $A_S/A_L \simeq 10$ of the two devices is comparable to the inverse of the corresponding length ratio $l_L/l_S = 14$, consistent with our suggested $A \propto 1/N (\propto 1/l)$ behavior. The resistance of the short CNFET is close to the quantum limit $\sim 6\,\mathrm{k\Omega}$ [31], indicating a transmission probability near unity and quasi-ballistic transport behavior for a channel length $\sim 500\,\mathrm{nm}$. On the other hand, the higher resistance of the long CNFET is a manifestation of its scattering-limited transport behavior. Taking into account this length dependence, the CNFET resistance can be expressed as

$$R = R_{\mathrm{SB}} + R_{\mathrm{diff}} = R_{\mathrm{SB}} + \left(\frac{h}{4e^2}\right)\frac{l}{\lambda}, \qquad (3)$$

where $\lambda$ is the electron mean-free-path in the carbon nanotube, and $R_{\mathrm{SB}}$ and $R_{\mathrm{diff}}$ are the resistance contributions due to the contact Schottky barriers and the scattering within the nanotube channel, respectively. Assuming the same $R_{\mathrm{SB}}$ for both devices[32], we estimate the scattering length $\lambda \sim 0.8\,\mu\mathrm{m}$ in the low-bias regime, in agreement with the values obtained from previous experiments[33] and theoretical predictions[34]. Although it has been suggested that the $1/f$ noise in bulk materials is induced by scattering with surface or/and bulk phonons[16] and the same may be true for nanotubes[35], the agreement between $A_S/A_L$ and $l_L/l_S$ is striking and suggests that $\alpha_H$ in CNFETs is not substantially influenced by the presence of acoustic phonon scattering in the long channel device. In addition, $\alpha_H$ is not affected by the ionized impurity-induced scattering, either. As shown in Fig. 3(b), the step-like features in the $I_d$-$V_g$ curves of the $n$-type CNFET is a manifestation of ionized impurity-induced ($\mathrm{K}^+$ ions) scattering in the nanotube channel[36]. However, the similarity of the gate-dependent $A$ exhibited by the undoped ($p$-type) and K-doped ($n$-type) CNFETs (Fig. 3) indicate little, if any, impact of the ionized impurity with regard to the $1/f$ noise in nanotubes. Since the low-frequency $1/f$ noise involves slow fluctuation processes with long time constants ($\geq \mu\mathrm{s}$), the electron-phonon or impurity scattering is, therefore, ineffective for the low-frequency $1/f$ noise phenomenon due to the much shorter scattering time in nanotubes ($\sim \mathrm{ps}$)[33, 37].

The relation of the noise amplitude $A$ and the sample resistance $R$ observed among various CNFETs can be consistently understood in terms of Eqs. (2) and (3). The two resistance components in Eq. (3), $R_{\mathrm{SB}}$ and $R_{\mathrm{diff}}$, exhibit dissimilar behavior with respect to the total carrier number $N$. While the resistance $R_{\mathrm{diff}}$ is always proportional to $N$ (because $l \propto N$), a low resistance $R_{\mathrm{SB}}$ is associated with a high transmission probability and is accompanied by a large $N$. For a short channel device where the transport is nearly ballistic, the total resistance $R \simeq R_{\text{SB}}$, and the noise amplitude $A$ exhibit a monotonic trend that increases with increasing $R$.

**CONCLUSION**

In summary, the $1/f$ noise behavior observed in semiconducting single-wall carbon nanotubes shows that the $A \propto 1/N$ relation is valid even for a 1D system with only tens to hundreds of carriers. While we have shown here the intrinsic noise strength ($\alpha_H$) in nanotubes is comparable to most bulk materials, the noise amplitude ($A$) can be quite significant because of the small number of carriers. While the $1/f$ noise phenomenon usually results from fluctuation processes with a wide distribution of time constants, the noise amplitude $A$ in carbon nanotubes is not affected by the nature of the transport behavior, i.e. quasi-ballistic or diffusive. This result holds important implications for technological applications not only based on nanotubes, but also for essentially any other nanostructures, e.g. nanowires, or aggressively scaled Si devices. On the other hand, the $N$-dependent noise in these nano-devices may also provide a novel technique to characterize the carrier numbers that are difficult to obtain otherwise.

**Acknowledgement**

The authors are indebted to J. Tersoff, R. Koch, and M. Freitag for their helpful discussions. We also thank B. Ek for expert technical assistance.

**Figure Caption**

Fig. 1: Schematics of various CNFET devices used to characterize the $1/f$ noise behavior in carbon nanotubes: (a) SB-CNFET, (b) C-CNFET, and (c) adjacent SB-CNFETs fabricated on the same nanotube with different channel lengths. The SB-CNFET is a back-gated device where the nanotube lies on a conducting substrate covered with a layer of oxide, and is contacted by two metal electrodes that serve as source and drain. In the C-CNFET, an additional middle gate stack, normally a metal electrode covered with a layer of oxide, is placed underneath the nanotube between the source and drain electrode.

Fig. 2: (a) Repeated (10 times) measurements of drain currents $I_d$ as a function of gate voltage $V_g$ of an SB-CNFET for different drain biases $V_d$. Inset: SEM image of the SB-CNFET. (b) The drain current as a function of time for various gate and drain voltages. The black, red, and blue traces corresponds to $V_d = -0.5\,\text{V}$, $-0.1\,\text{V}$, and $-0.01\,\text{V}$, respectively. (c) Normalized current noise power spectrum ($S_I/I^2$) of the CNFET at various gate voltages. The drain bias is kept at $10\,\text{mV}$. The solid line indicates the $1/f$ dependence of the noise power spectrum.

Fig. 3: (a) The drain current (left axis) and $1/f$ noise amplitude $A$ (right axis) as a function of $V_g$ of an SB-CNFET. The solid and dash lines are simulation results assuming an $1/f$ noise parameter $\alpha_H = 2 \times 10^{-3}$. (b) The drain current (left axis) and $1/f$ noise amplitude $A$ (right axis) as a function of $V_g$ of an $n$-type SB-CNFET after being exposed to potassium vapor. The drain bias is $10\,\text{mV}$ and the solid line is a guide to the eye.

Fig. 4: (a) SEM of a dual-gate CNFET, showing the $Al_2O_3/Al$ gate stack underneath the nanotube and between the two contacts. (b) The drain current (left axis) and $1/f$ noise amplitude $A$ (right axis) as a function of Al gate voltage $V_{g-Al}$ of the dual-gate CNFET for two Si backgate voltages $V_{g-Si}$ =-2.5 V and -1.5 V. The drain voltage is $10\,\text{mV}$.

Fig. 5: Normalized current noise power spectrum ($S_I/I^2$) of two CNFETs with channel lengths of $7\,\mu\text{m}$ ($L$) and $500\,\text{nm}$ ($S$). The noise amplitudes $A$ of the long and short devices are $4 \times 10^{-7}$ and $4 \times 10^{-6}$, respectively. Inset: SEM image of two CNFETs, denoted by $S$ and $L$, fabricated using the same nanotube. The middle electrode is shared by the two CNFETs as the source electrode.

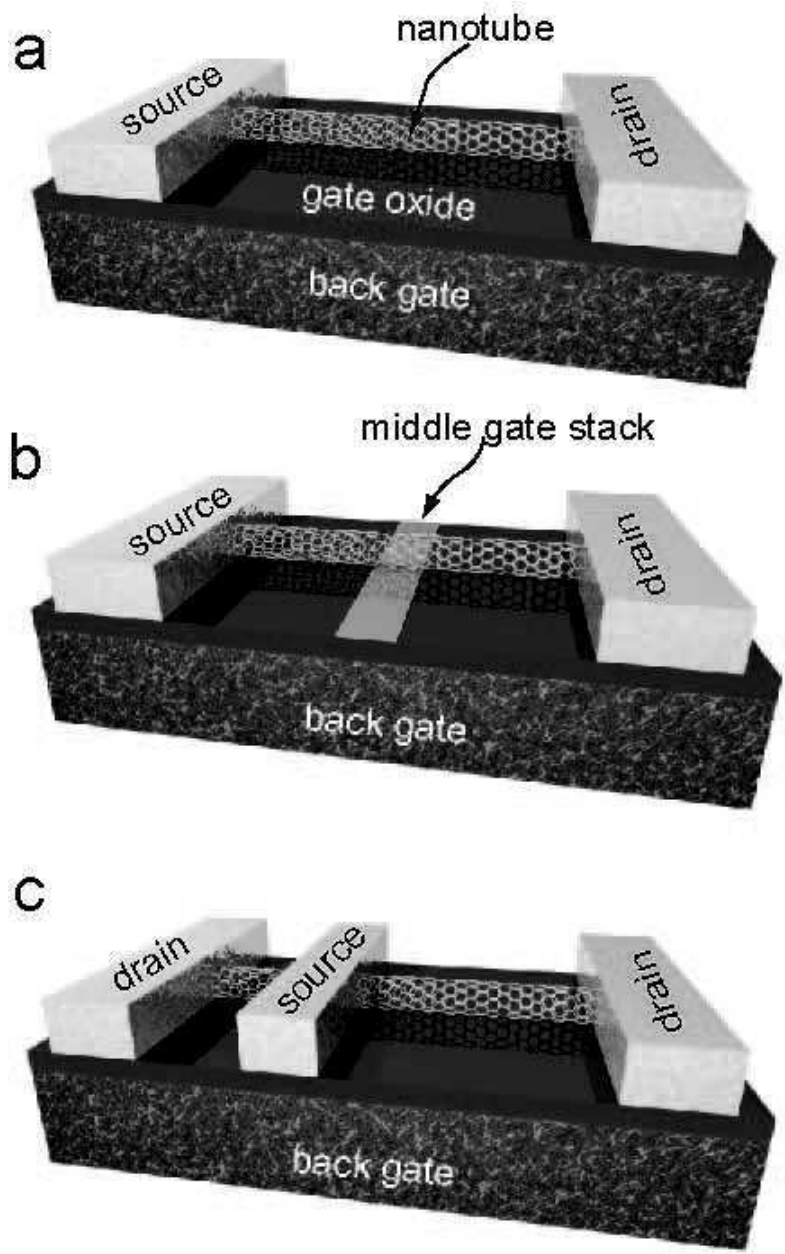

FIG. 1: Lin et al.

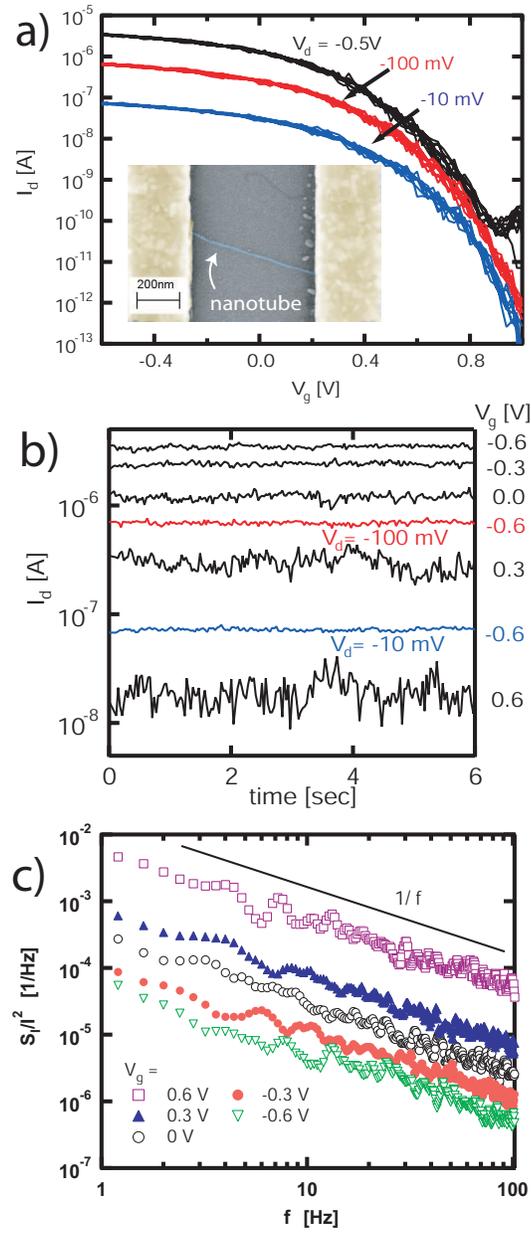

FIG. 2: Lin et al.

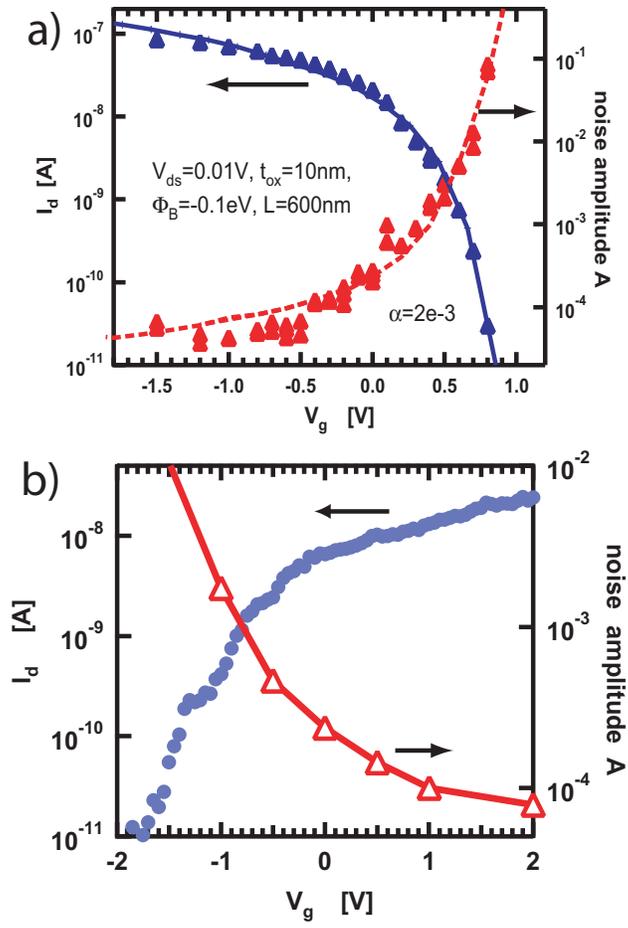

FIG. 3: Lin et al.

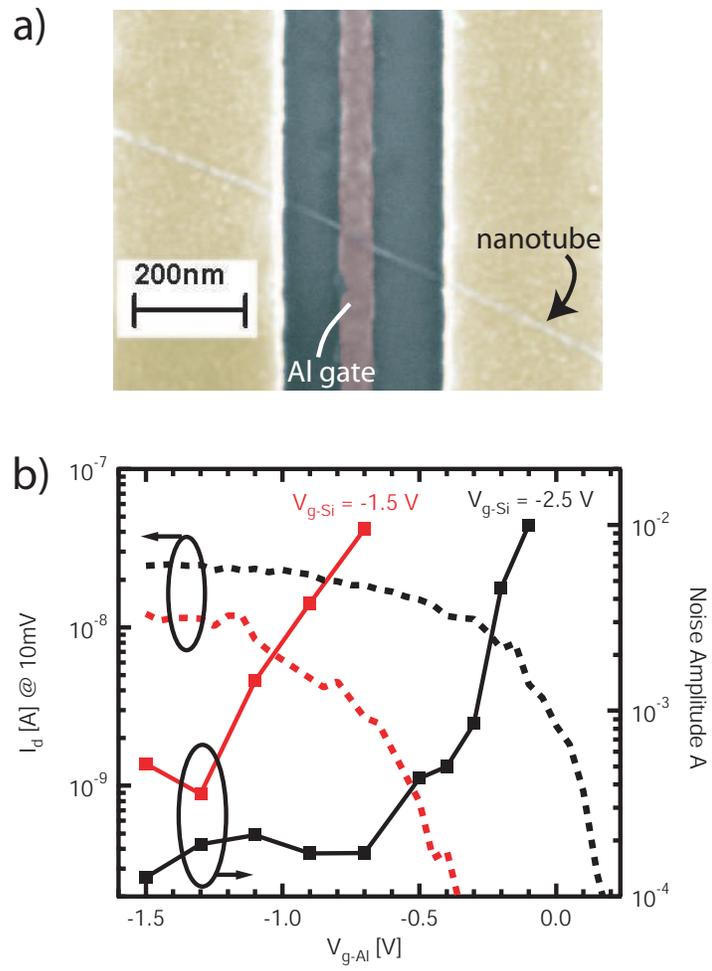

FIG. 4: Lin et al.

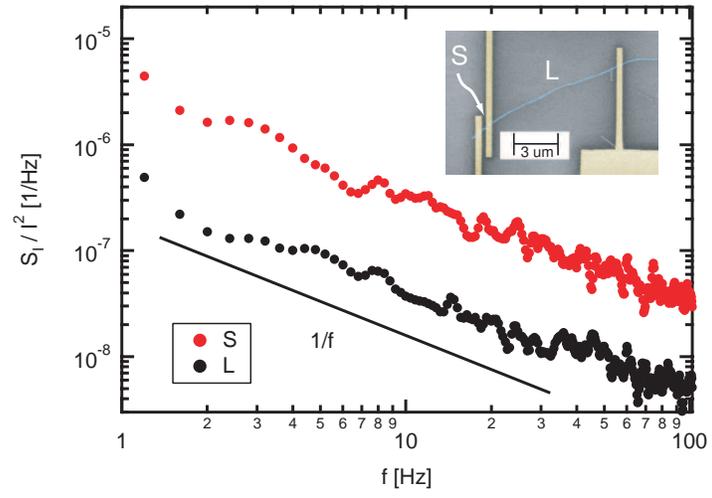

FIG. 5: Lin et al.